\documentstyle [preprint,aps]{revtex}
\begin{document}
\title{Bose-Enhancement and Pauli-Blocking Effects in Transport Models}
\author{Scott Pratt and Wolfgang Bauer}
\address{National Superconducting Cyclotron Laboratory\\
and Department of Physics and Astronomy\\
Michigan State University, East Lansing MI 48824-1321, USA}
\maketitle

\begin{abstract}
The ability of semi-classical transport models to correctly simulate
Pauli blocking and Bose enhancement is discussed.  In the context of
simple quantum mechanical systems, it is shown that using $(1 \pm f)$
enhancements is inadequate to describe systems far from equilibrium.
The relaxation of  $(1 + f)$ descriptions toward equilibrium is
studied both in the context of simple models, including a hot pion gas
confined to a box.  We give simple estimates for the characteristic
number of collisions for a system to relax toward equilibrium.
\end{abstract}

\vspace*{14pt}

Semi-classical simulations form the backbone of the phenomenology for
high-energy and intermediate-energy heavy-ion physics. The
semi-classical nature of such descriptions is justified if  Compton
wave lengths are smaller than the size of the reaction volume, and if
phase space filling factors, $f$, are much less than unity.  The first
condition is generally fulfilled whenever incident beam energies
exceed $\approx 40\cdot A$ MeV.  The second condition is violated for
lower energy collisions by nucleons, as excitation energies are less
than the Fermi energy until beam energies exceed $\approx 100\cdot A$
MeV. For relativistic heavy-ion collisions, $E/A > 10$ GeV,
phase-space-filling factors for pions also approach unity.  For these
two regimes, quantum statistics must be incorporated into simulations.

Nordheim \cite{Nordheim} first pointed out that modifying the
Boltzmann transport equation by incorporating phase-space occupation
factors $(1 \pm f)$ into the collision integral, showing the phase
space distribution functions have the Bose-Einstein or Fermi-Dirac
distributions as their respective equilibrium solutions.  The usual
way to accomplish this is by modifying the cross sections from their
in-vacuum values $\sigma_{ab\rightarrow cd}^{vac}$ to the in-medium
value\cite{BKD84,ZL68,WLB91,WB92}:

\begin{equation}
\sigma^{med}_{ab\rightarrow cd} =
\sigma^{vac}_{ab\rightarrow cd} \cdot (1 \pm f_c) (1 \pm f_d).
\end{equation}

Given a system with measured occupations, a scattering will choose its
final state according to the weights described above.  This problem
was originally cast in terms of the Einstein $A_{nm}$ and $B_{nm}$
coefficients\cite{Feynman}, which is the correct treatment given
measurements are made after each emission. However, when measurements
are not made between individual scatterings; correct treatment of Bose
and Fermi statistics requires the incorporation of many-body effects.
Our principal objective here is to compare models where full $n$-body
symmetrization has been accounted for with the $(1\pm f)$ method.

We start by examining the Einstein problem:  If in free space the rate
at which quanta are scattered into a cell $k$ is $\alpha_k$ and the
rate at which quanta are scattered out of the cell is $\beta_k f_k$,
the rate equations for the phase-space occupation $f_k$ in and out of
the medium are given by:

\begin{eqnarray}
\frac{df_k^{vac}}{dt} &=& \alpha_k - f_k^{vac} \beta_k.\\
\frac{df_k^{med}}{dt} &=& \alpha_k (1 \pm f_k^{med})
                        - f_k^{med} \beta_k.
\end{eqnarray}
This results in equilibrium values of $f_k$,

\begin{eqnarray}
f_k^{vac} &=& \frac{\alpha_k}{\beta_k}\\
f_k^{med} &=& \frac{\frac{\alpha_k}{\beta_k}}
                   {1\mp\frac{\alpha_k}{\beta_k}}.
\end{eqnarray}

If the free space values of $\alpha_k$ and $\beta_k$ result in a
Boltzmann factor, $\alpha_k / \beta_k = \exp(-(E_k-\mu)/T)$, the
in-medium result is a Bose-Einstein or Fermi-Dirac form.  Thus one can
be confident that a simulation that incorporates $(1 \pm f)$
corrections will ultimately obtain correct equilibrium behavior given
sufficient time such that gain and loss terms balance.

In a relativistic heavy-ion collision ($E/A \approx 200$\ GeV) the
reaction can be viewed as having two stages.  During the creation
stage pion production takes place, while during the reinteraction
stage expansion and cooling occur, during which the pion number stays
roughly constant.  We divide our model into two stages
correspondingly.  For the first stage we consider only production.
Given the large energies of the colliding nucleons, reformation of the
energetic initial states is negligible and one can consider only
formation processes.  In the absence of Bose effects, we assume the
probability of having created $n$ pions is Poissonian:

\begin{equation}
P_k^0(n,t) = \exp(-\eta_k(t)) \frac{\eta_k(t)^{n}}{n!}.
\end{equation}

Fully including symmetrization, the probability of creating $n$ pions
is enhanced by a factor of $n!$ assuming the time $t$ is after
emission has stopped and that no measurements were made at
intermediate times.  This enhancement can be considered as either the
contribution of interference terms or as the amplitude of a matrix
element with $n$ identical Boson creation operators, $\langle 0|a^n
a^{\dagger n}|0\rangle$.

\begin{eqnarray}
\label{nfactenhance}
P_k^{full}(n,t) &=& \frac{1}{Z_k} P_k^0(n,t) n!\\
\nonumber
&=& (1-\eta_k(t))\eta_k(t)^n
\end{eqnarray}
where $Z_k$ is a normalization.

If the state is measured at each intermediate time $t$ during the
emission, $(1 \pm f)$ methods are justified.  The probability of
seeing $n$ particles at time $t$, $P_k^{(1+f)}(n,t)$ is then
determined by the equation:

\begin{equation}
\label{timedev}
\frac{d}{dt}P_k^{(1+f)}(n,t) = \frac{d\eta_k(t) }{dt}
\left\{ n\cdot P_k^{(1+f)}(n,t) - (1+n)\cdot P_k^{(1+f)}(n+1,t) \right\}
\end{equation}

\noindent
with the solution

\begin{eqnarray}
P_k^{(1+f)}(n,t) &=& (1 - \gamma_k (t))\gamma_k(t)^n\\
\label{1+fsol}
\nonumber
\gamma_k(t) &=& 1-\exp(-\eta_k(t))
\end{eqnarray}

Simulating systems populated according to $(1+f)$ weights results in
dendritic growth of phase-space occupation.  The first few particles
can bias the behavior of all the following particles, leading to large
fluctuations in the populations.  Whereas, when no measurement is made
until the end of the emission, all of the emissions are affected
equally by the symmetrization. The average occupation is lower for the
$(1+f)$ case, Eq. (\ref {timedev}), than for the case where full
$n$-body symmetrization weighting is applied at the end of the
emission as in Eq. (\ref {nfactenhance}), but certainly higher than
for the case where symmetrization is neglected, $\langle n_k^0
\rangle$.

\begin{eqnarray}
\langle n_k^0\rangle &=& \eta_k(t\!=\!\infty )\nonumber\\
\label {nbar}
\langle n_k^{(1+f)}\rangle &=&
  \frac{\gamma_k(t\!=\!\infty )}{1-\gamma_k(t\!=\!\infty )}\\
\nonumber
\langle n_k^{full}\rangle &=&
  \frac{\eta_k(t\!=\!\infty )}{1-\eta_k(t\!=\!\infty )}
\end{eqnarray}

If the function $\eta_k$ has a thermal form, $\eta_k =
\exp(-(E_k-\mu)/T)$, the full $n$-body result corresponds to the
thermal Bose-Einstein form.  The additional created particles account
for the difference of the Bose-Einstein form as compared to the
Boltzmann form for the spectra. For the $(1+f)$ case, the population
of the different levels of energy $E_k$ can not be described by one
temperature and chemical potential.

Fig.\ \ref{fvseta}(a) illustrates the three solutions for the mean
occupancy. There is more than just a quantitative difference.  The
$(1+f)$ result never diverges as compared to the full $n$-body case
which results in an infinite occupancy when $\eta_k$ exceeds unity.
The $(1+f)$ result even differs to the second power of $\eta_k$.

One can solve the same models for fermions.  In this case $(1+f)$
terms are replaced by $(1-f)$ terms in Eq. (\ref{timedev}) and the
$n!$ in Eq. (\ref {nfactenhance}) is replace by zero for $n>1$.  The
results look very similar for the mean occupations.

\begin{eqnarray}
\langle n_k^{(1-f)}\rangle &=& \gamma_k(t=\infty )\\
\nonumber
\langle n_k^{full}\rangle  &=&
\frac{\eta_k(t=\infty )}{1+\eta_k(t=\infty )}
\end{eqnarray}
The difference is illustrated in Fig. \ref{fvseta}(b) which shows that
$(1-f)$ models can modestly underestimate Pauli-blocking effects.

Even if $(1+f)$ models grossly underestimate initial populations of
low-energy states, rescattering effects can lead to equilibrium.  If
sufficient scattering occurs to produce equilibrium, the only fault of
the $(1+f)$ models will lie in the prediction of the overall
multiplicity.  For this reason we wish to investigate the amount of
scattering required to produce equilibrium.

We consider a closed system with $N_l$ levels labeled by $k$ where
each level has a degeneracy $d_k$ and base weights $\eta_k$.  For
initial conditions we fill the levels with $N$ particles according to
the $(1+f)$ prescription, neglecting any loss terms.  Collisions are
modeled by randomly removing one of the $N$ particles and placing it
into a level according to the weight $d_k (1+n_k) \eta_k$.  After
sufficient collisions populations approach equilibrium.

The equilibrium population for fixed $N$ systems can be calculated by
using diagrammatic methods\cite{Pratt93}.  Although these method have
only been used to study multi-particle symmetrization of outgoing
states, they can also be used to calculate equilibrium populations.
Ref. \cite{Pratt93} describes how to calculate the probability of
emitting $N$ particles in states $k_1 \cdots k_N$ given the source
function $S(p,x)$, where $S(p,x)$ is the probability of emitting a
particle from space-time point $x$ into final state $p$ neglecting
symmetrization.  Under the assumption that the momentum dependence in
$S$ is not correlated to the space-time dependence, $S(p,x) = S(x)
\eta (p)$ the expression in Ref. \cite{Pratt93} becomes:

\begin{equation}
P(k_1,\cdots k_N) = \frac{1}{Z} \int dx_1 \cdots dx_N
\eta_{k_1} \cdots \eta_{k_N} S(x_1) \cdots S(x_N)
|U(x_1,\cdots x_n;k_1,\cdots k_N)|^2,
\end{equation}
where $U$ is the evolution matrix from the creation of pions at $x$ to
their measurement in states $k$.  If we assume that the system is
finite with discreet eigenstates, and if we assume that there is no
${\bf x}$ dependence aside from confining the region of emission, the
expression above becomes:

\begin{equation}
\label{p(k)}
P(k_1,\cdots k_N) = \frac{1}{Z} \int dx_1 \cdots dx_N\,
\eta_{k_1} \cdots \eta_{k_N}
|\langle x_1,\cdots x_n|k_1,\cdots k_N\rangle |^2.
\end{equation}

The squared quantity is the squared n-particle wave function with $N!$
terms. The same equation applies for an open (non-bounded) system if
the emission occurs all at one time, except that the labels refer to
continuum states rather than discreet eigenstates.  By making the
substitution $\eta_i = \exp(-E_i/T)$ one sees that Eq. (\ref{p(k)}) is
equivalent to a thermal trace, hence it is the equilibrium answer.

Methods for accounting for all $N!$ terms in the outgoing wave
function were devised in Ref.\ \cite{Pratt93}.  Translating the method
of Ref.\ \cite{Pratt93} to the case of discreet states results in the
prescription below for calculating $\langle n_k \rangle$ given
$\eta_k$ and $N$.  First define quantities $G_n(k)$, $C_n$ and
$\omega(n)$.

\begin{eqnarray}
G_n(k) &\equiv & (\pm 1)^{n-1} d_k \eta_k ^n\\
\nonumber
C_n(k) &\equiv & \frac{1}{n}\sum_k G_n(k)\\
\nonumber
\omega(n) &\equiv & \sum_{i_1\cdots i_n}
\prod \frac{C_1^{i_1}}{i_1!}\frac{C_2^{i_2}}{i_2!}
\cdots \frac{C_n^{i_n}}{i_n!}
\end{eqnarray}

The sum over $i_1 \cdots i_N$ is constrained such that the overall
order is $N$, $i_1 + 2i_2 + 3i_3 +\cdots = N$. An algorithm for
calculating the sum in the expression of $\omega (n)$ is also given in
Ref.\ \cite{Pratt93}.  The expression for the equilibrium occupations
is:

\begin{equation}
\langle n_k\rangle _{equil} = \frac{G_1(k)\omega (n-1)
     + G_2(k)\omega (n-2)
+ \cdots G_{n-1}(k)\omega (1) + G_n(k)}{\omega (N)}.
\label{equil}
\end{equation}

Now we return to the evolution of $\langle n_k\rangle $ averaged over
many samplings.  Fig.\ \ref{evolve}(a) shows the occupation as a
function of the number of collisions for the case where there are 8
levels with filling weights $\eta_i = \exp(-k\cdot \Delta E/T)$, with
$\Delta E = T/8$, and 40 particles.  Given the large number of
particles in a limited phase space one can easily see that the initial
conditions underpopulate the ground state, but that a sufficient
number of collisions restores equilibrium.  The dashed lines in Fig.\
\ref{evolve}(a) represent the equilibrium occupancies calculated
above.  In order to make a detailed study of the approach to
equilibrium Fig. \ref{evolve}(b) shows the discrepancies from
equilibrium as a function of the number of collisions on a logarithmic
scale.  This allows the extraction of characteristic scales for the
exponential approach to equilibrium.

One striking aspect of Fig. \ref{evolve}(b) is that as one approaches
equilibrium the lines become parallel.  The inverse slope of this line
corresponds to the characteristic number of collisions for approaching
equilibrium $\tau_d$.  In order to understand how the parameters
(level density, temperature, ground-state population) determine
$\tau_d$, we linearize the evolution equation and treat $f_k$ as a
continuous variable.  This allows us to quickly find $\tau_d$ for a
variety of systems, which will lend insight into what characteristics
determine $\tau_d$.  Ignoring the discreet nature of $f_k$,

\begin{eqnarray}
\frac{df_k}{d\tau} &=& \frac{(1+f_k) \eta_k }{Z} - \frac{f_k}{N}.\\
\nonumber
Z &=& N \sum (1+f_k) \eta_k.
\end{eqnarray}

The normalization factor $Z$ which is a function of the distribution
$f_k$, ensures that particle number is conserved.  For small
deviations from equilibrium, one can expand $Z$ keeping the lowest
order terms in $\delta f_k = f_k - f_k^{eq}$ to obtain:

\begin{equation}
\frac{d \delta f_k}{d\tau} = \frac{- \delta f_k}{1+f^{eq}_k}
+\frac{f^{eq}_k}{N} \sum_j \frac{\delta f_i}{1+f^{eq}_j}.
\end{equation}

One can solve the $N$ coupled differential equations with matrix
methods.  There are $N-1$ solutions given the constraint $\sum_k
\delta f_k =0$.  Each solution has the form:

\begin{equation}
\delta f_k \propto \exp \left\{ -\frac{N_{coll}}{N\tau_d}\right\}.
\end{equation}

Each eigenvector has a different value of $\tau_d$.  We find  that the
slowest decaying eigenvector is the one where the ground state is
furthest from equilibrium.  For thermal systems we also find one can
estimate $\tau_d \approx T/(E_2-E_1)$ where $E_1$ and $E_2$ refer to
the energies of the ground state and the first excited state.  This
approximation  works to better than 20\% accuracy whenever the first
state has an occupation much greater than unity.  This is true even
when the second state is degenerate.  Thus in the case of a nearly
degenerate ground state, equilibration is very slow.

The simulation of the eight-level model described above can be easily
modified to handle more sophisticated systems.  To that end we
consider the eigenstates of a cube of length $L$ = 10 fm on a side
with a base filling factor, $\eta_i = exp(-E_i/T)$, with $T$ = 175
MeV.  Assuming the eigenstates are determined by a relativistic
dispersion:

\begin{equation}
E^2_{k_x,k_y,k_z} = m^2 + \frac{\pi^2}{L^2}
\left( k_x^2 +k_y^2 +k_z^2 \right),
\end{equation}
we simulate the behavior of $\langle f_i\rangle $ for 200 particles.
Fig. \ref{fbox} shows symmetrization enhancements to spectra.  The
symmetrization enhancement is defined by the average phase-space
filling for states of that energy divided by $\exp(-E_i/T)$. The
lowest curve demonstrates the symmetrization enhancement after states
have been filled according to the $(1+f)$ prescription with no
collisions.  The figure shows that after five collisions per particle
symmetrization enhancements for small energies are half way to
equilibrium values.  The estimate of $\tau_d$ mentioned above for this
system is six collisions per particle.

In heavy-ion collisions the number of collisions pions feel is on the
order of a half dozen for sulphur on lead collisions at 200 GeV
incident laboratory energy. For the upcoming lead beam that number
could roughly double.  Thus we see that for larger systems, $(1+f)$
theory should be able to recover from its shortcomings and predict a
spectra in line with what would be expected had all $n$-body effects
been included.  However, the multiplicity would still be
underestimated.  The amount by which the multiplicity might be
underestimated would be similar to the amount of emission due to
stimulated emission.  Fig. \ref{fvseta} shows that the extra
multiplicity could possibly be very large if values of $\eta_k$ become
larger than unity.  Although a real system can not emit an infinite
number of particles which is possible with the Poissonian model used
above, the possibility of a significant fraction of pions being due to
enhanced emission is possible.  Phase-space occupations should be even
more overpopulated for conditions created by the relativistic heavy
ion collider (RHIC), scheduled to come on line in 1999.

In proton-induced collisions, conclusions are simpler.  For these
systems reinteraction of hadrons is not so important.  Here $(1+f)$
models are completely inappropriate and full $n$-particle
symmetrization effects must be incorporated.  This issue has been
addressed in the context of {\sc centauro} phenomena
\cite{Lam,Pratt93,Zel94,Lattes}.

The possibility of creating super-radiant pion sources is one of the
most exciting prospects of relativistic-heavy-ion physics, cosmic-ray
physics and  particle physics.  Symmetrization phenomena should
manifest themselves in a variety of observables: multiplicity
distributions, isospin distributions and spectra. Making confident
predictions of collective symmetrization phenomena is inherently
difficult due to the unstable nature of enhanced emission.  We have
shown here that $(1+f)$ based simulations can grossly underpredict
such phenomena.  Should super-radiant conditions be attained in
relativistic heavy-ion collisions, one of the most daunting
theoretical challenges will be finding a way to incorporate many-body
symmetrization into descriptions where emission and collisions are
treated realistically.

{\acknowledgements  {This work was supported by a Presidential Faculty
Fellow Award, NSF grant no.\ PHY-925355, and by NSF grant no.\
PHY-9017077.}}

\begin{figure}
\caption[]{a: Average population of a Bosonic state as a function of
the Poissonian parameter $\eta$ for the cases of neglected
symmetrization (dotted line), $(1+f)$-enhancement (dashed line), and
full $n$-body enhancement (solid line). b: Same comparison for a Fermi
gas; $(1-f)$ suppression (dashed line), full $n$-body suppression
(solid line).}
\label{fvseta}
\end{figure}

\begin{figure}
\caption[]{Eight level system with 40 particles and level spacing
$\Delta E = T/8$. \\ a: Average populations of the eight levels as a
function of the number of collisions per particle.  Dashed lines:
exact equilibrium values computed from Eq.\ \protect{\ref{equil}}.
Solid lines: $(1+f)$-type simulations, averaged over $3\times 10^6$
events.\\ b: Deviation of the $(1+f)$-type simulations from
equilibrium populations.  After many collisions the lines become
parallel.  Statistical noise causes the ragged behavior for small
deviations.}
\label{evolve}
\end{figure}

\begin{figure}
\caption[]{Populations for states, divided by $\exp(-E_i/T)$, as a
function of the energy for a system of 200 relativistic particles
confined to a 10 fm cube at a temperature of 175 MeV, obtained from
$(1+f)$-type simulations and averaged over $10^5$ events.  Initial
populations (circles), after one (crosses), two (diamonds), five
(squares), and fifty (upright crosses) collisions per particle.  After
many collisions populations have approached equilibrium values (dashed
line).}
\label{fbox}
\end{figure}

\end{document}